\documentclass[10pt, preprint]{aastex}

\usepackage{graphicx}
\usepackage{epsfig}

\begin{document}
\title{SUB-WAVELENGTH RESOLUTION IMAGING OF THE SOLAR DEEP INTERIOR}
\author{Shravan M. Hanasoge$^1$}
\affil{W. W. Hansen Experimental Physics Laboratory, Stanford University, Stanford, CA 94305}
\altaffiltext{1}{Now at Max-Planck-Institut-f\"{u}r-Sonnensystemforschung, 37191 Katlenburg-Lindau, Germany}
\author{Thomas L. Duvall, Jr.}
\affil{Solar Physics Laboratory, NASA/Goddard Space Flight Center, Greenbelt, MD 20771}

\email{shravan@stanford.edu}
\begin{abstract}
We derive expectations for signatures in the measured travel times of waves that interact with 
thermal anomalies and jets. A series of numerical experiments that involve the dynamic
linear evolution of an acoustic wave field in a solar-like stratified spherical shell in the presence of
fully 3D time-stationary perturbations are performed. The imprints of these interactions are
observed as shifts in wave travel times, which are  extracted from these data through methods of
time-distance helioseismology \citep{duvall}. In situations where at least one of the spatial 
dimensions of the scatterer was smaller than a wavelength,
oscillatory time shift signals were recovered from the analyses, pointing directly to a means of 
resolving sub-wavelength features. As evidence for this claim, we present analyses of simulations with
spatially localized jets and sound-speed perturbations. We analyze 1 years' worth solar observations to 
estimate the noise level associated with the time differences. Based on theoretical estimates, Fresnel zone time shifts
associated with the (possible) sharp rotation gradient at the base of the convection zone are of the order 0.01 - 0.1 s,
well below the noise level that could be reached with the currently available amount of data ($\sim 0.15-0.2$ s with 10 yrs of data). 

\end{abstract}
\keywords{Sun: helioseismology---Sun: interior---Sun: oscillations---waves---hydrodynamics}

\section{INTRODUCTION}
The deep interior of the Sun holds many a secret in its opaque clutch. Our current understanding of the
properties of these regions comes predominantly from the application of 
methods of global helioseismology to observations of solar oscillations \citep[for reviews, see e.g.][]{jcd02,jcd03}. 
However, many aspects of the internal structure and dynamics of the Sun are non-global in nature
and may therefore benefit from investigations via techniques of local helioseismology \citep[see review by][]{gizon05}.
Small persistent jets and thermal anomalies in the convection zone, radial variations in the rotation rate at the 
bottom of the convection zone, convective flows in
the interior etc. are examples of such spatially localized phenomena.

A stumbling block associated with seismic studies of the deep interior is the substantial 
wavelength that propagating $p$ modes attain in this region.
At the bottom of the convection zone, the acoustic wavelength 
is approximately 30 times larger than at the photosphere; a 4 mHz wave with wavelength 2 Mm
at the photosphere attains a size of 58 Mm, $\sim 8$\% of the solar radius by the time it reaches
said bottom. The classical imaging resolution limit of waves with wavelength $\lambda$ is $\lambda/2$, and in
practice, more like $\lambda$. This places severe restrictions on our ability to infer properties of
the deep interior of the Sun. In the field of optical and acoustic tomography, this resolution limit
has been overcome by recognizing that present in the near field of a scatterer is a collection
of evanescent waves that contain information about its sub-wavelength structure \citep[e.g.][]{maynard,boz}. That this 
principle has an analog in the solar case has been theoretically discussed \citep{bogdan,hanasoge07b} but
conclusive observations of this envelope of evanescent modes are still lacking. And since we are unable to 
directly observe the bottom of the convection zone, there seems little hope in being able to utilize 
the near-field signal. Moreover it is not clear how significant a role these evanescent waves play.

The interactions of $p$ modes with localized spatial anomalies in the backdrop
of a solar-like stratified medium and the resultant changes in wave travel times are well understood in a range of situations 
\citep[e.g.][]{birch00,gizon02}. Theory and observation have shown that sub-wavelength sized scatterers
can cause significant and observable oscillations in the time shifts \citep{duvall06} and global mode frequencies \citep[e.g.][]{jcd95}. Moreover, from the work of
\citet{birch00}, it is clear that the stratification is one of the causes of complex interference patterns that contributes to
the oscillations (Fresnel zones) in the travel times. Another participating factor
is the limited chromatic extent (25 - 40\% of a decade) of the solar wave spectrum; highly bandlimited wave packets
are known to produce striking interference patterns because of the emergence of a 
unique interference length scale (the wavelength). 

Wave interactions in local helioseismology are primarily characterized using 
approximations in the ray, Rytov, or Born limits. One situation wherein the ray approximation is 
accurate is when the wavelength
is much smaller than the characteristic spatial size of the perturbation. This requirement invalidates
the ray approximation for a large fraction of deep interior studies.
As for the Born approximation, \citet{gizon06} have theoretically shown 
in the context of thin flux tubes that it may break down in the limit of vanishing
flux tube radius to imaging wavelength ratio. Thus, extremely small scatterers (at least in the case
of magnetic fields and possibly other perturbations) are not well described by the Born limit. Therefore, 
imaging sub-wavelength aspects of the deep interior using interpretations derived from the Born approximation
may also not be very accurate. 

Numerical simulations of the solar wave field in full 3D spherical geometry \citep{hanasoge1} 
provide a means of addressing the wide variety of interaction phenomena described above in a consistent manner.
The methods of realization noise subtraction \citep{hanasoge07a} and deep-focusing time-distance
helioseismology \citep{duvall03} are applied in the analysis of the simulation data. In $\S$\ref{sims.descrip}, we discuss
the numerical methodology and the sound-speed perturbations studied here. We discuss the deep-focusing geometry and
time-distance methods and present results from these calculations.  The appearance of Fresnel zones in analyses of simulations containing
jets at the base of the convection zone is discussed in $\S$\ref{jets}. From time-distance analyses of observations, we attempt in $\S$\ref{observations} 
to search for oscillatory time shifts arising as a consequence of the possibly sharp rotation gradient at the base of the convection zone. 
Finally, we summarize and conclude in $\S$\ref{conclude.sec}.

\section{Simulations and test cases}\label{sims.descrip}
The numerical code developed in \citet{hanasoge1} and \citet{hanasoge.phd} is the starting point for the 
results presented herein. The linearized Euler equations in spherical geometry are spatio-temporally evolved
in a spherical shell extending from $r = 0.24 R_\odot$ to $1.002 R_\odot$, where $R_\odot$ is the radius
of the Sun. Spatial derivatives are calculated using spherical harmonic representations in the horizontal 
directions while sixth-order accurate compact finite differences \citep{lele} are implemented in the radial 
direction. Temporal evolution is achieved through the repeated application of an optimized second-order
accurate Runge-Kutta scheme \citep{hu}. The boundaries are rendered absorbent through the application of
boundary conditions prescribed by \citet{thompson} in conjunction with damping sponges placed adjacent to the
lower and upper radial boundaries. Waves are excited by a phenomenological forcing term in the radial 
momentum equation (dipolar sources). The source function is computed in spectral space ($l,m,\omega$) as a
set of Gaussian-distributed random numbers for each Fourier component, multiplied by a frequency envelope so
as to mimic the solar-like wave power distribution in frequency. Note that $(l,m)$ are the spherical
harmonic indices while $\omega$ is the angular frequency. The radial component of the oscillation
velocity is extracted at a height of 200 km above the photosphere at each minute. Since the near-surface layers
of the Sun are highly convectively unstable, we use the artificially stabilized version described in
\citet{hanasoge.phd} (and hence different from standard models of the Sun, e.g. Christensen-Dalsgaard et al. 1996).

\begin{figure}[!ht]
\centering
\epsscale{1.}
\plotone{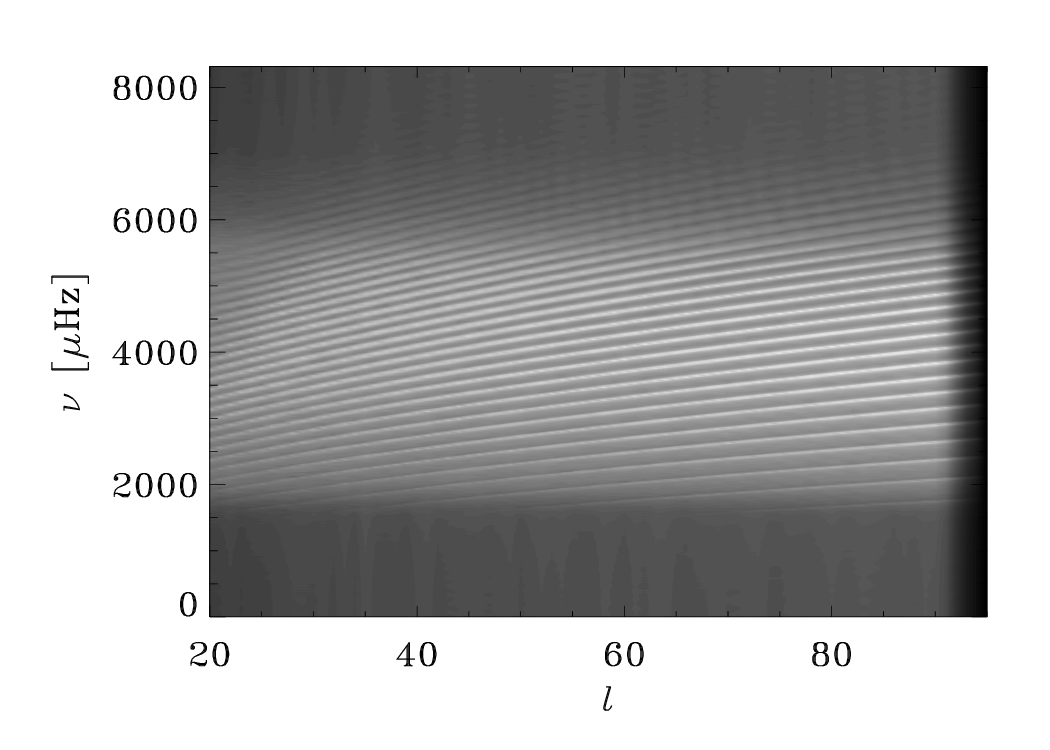}
\caption{Modal power distribution as a function of frequency ($\nu$) and spherical harmonic degree ($l$) from
a 24-hour long simulation with $l_{max} = 95$. Standard line of sight Doppler signals are substitued by the radial component 
of the oscillation velocities, which are extracted at an altitude of 200 km above the photosphere at a cadence 
of once per minute. The $l \lesssim 20$ modes propagate in proximity of the lower boundary, and are damped 
out by the damping sponge described in $\S$\ref{sims.descrip}, hence not shown here. }
\label{power.spectrum}
\end{figure}

\begin{figure}[!ht]
\centering
\epsscale{0.75}
\plotone{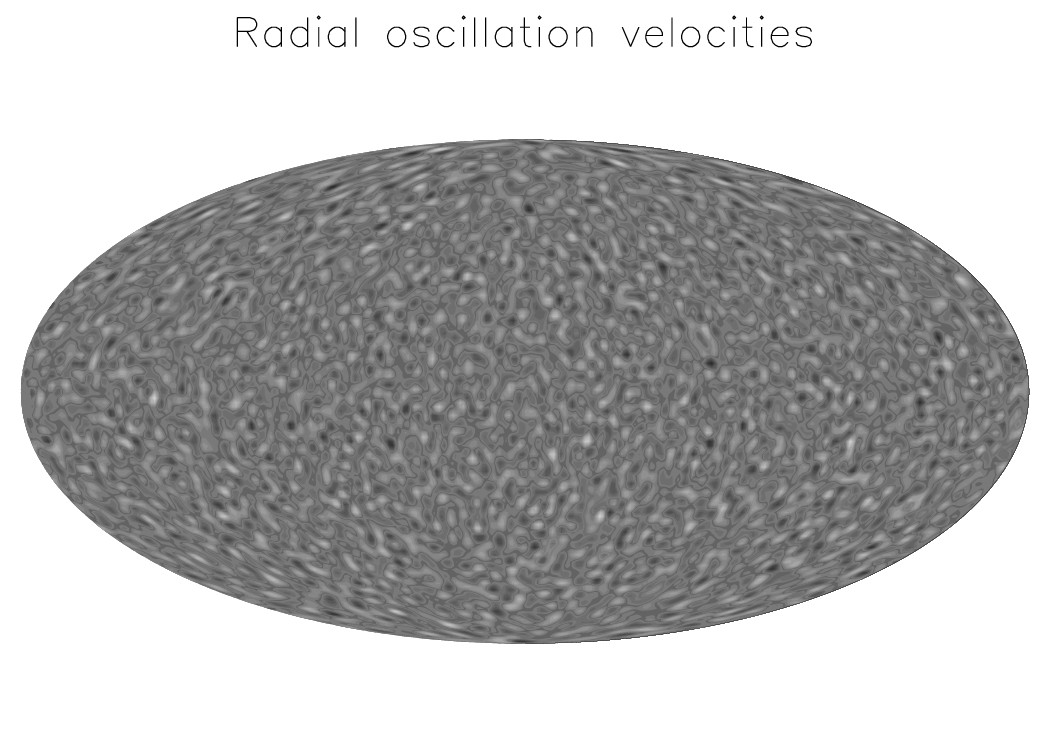}
\epsscale{0.75}
\plotone{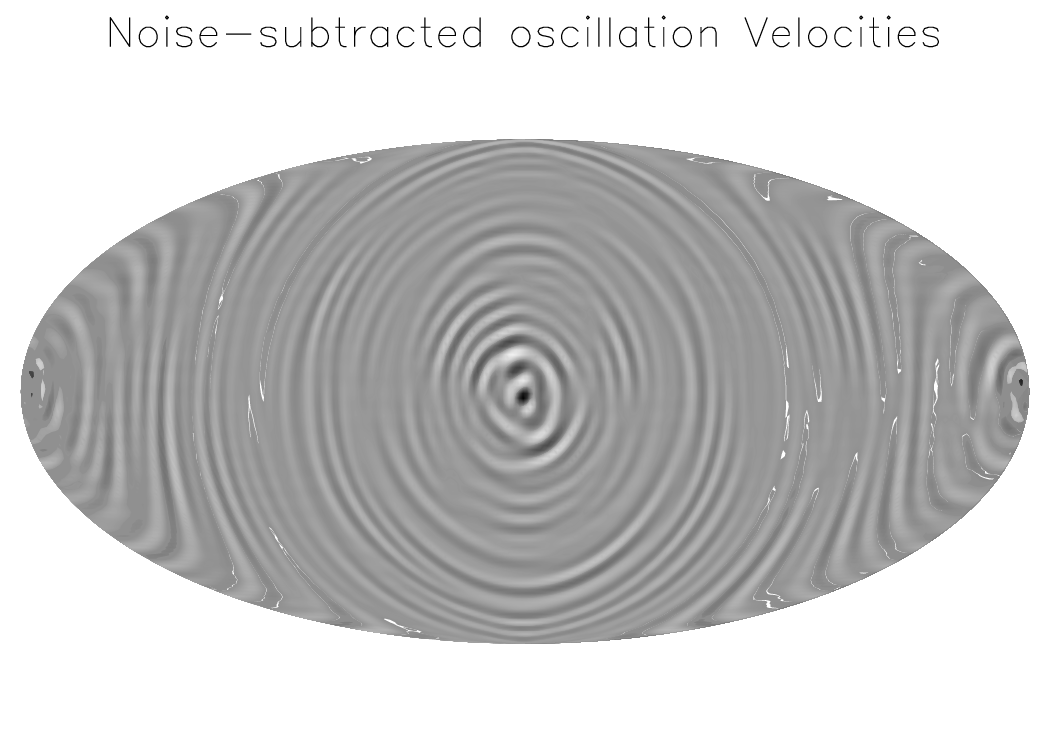}
\caption{Mollweide projection of the radial oscillation velocities derived from a simulation with a sound-speed perturbation -
displayed are raw oscillation data (upper panel) and noise-subtracted velocities (lower panel). The scattering, invisible in 
the upper panel, becomes far more evident when the realization noise subtraction is performed, i.e. when the velocities
of the quiet data are subtracted. This principle also applies to the travel times. Note that the method of noise subtraction
is entirely theoretical; no analog exists when dealing with real observations.}
\label{noise.sub}
\end{figure}

An example power spectrum from a simulation is displayed in Figure~\ref{power.spectrum}. Because we do not 
incorporate realistic damping, the power distribution as a function of frequency peaks at approximately 4 mHz, higher
than the solar power peak, which occurs at 3.5 mHz. A consequence of this is that the wave packets which contribute to 
the analyses have systematically shorter wavelengths in comparison to the Sun. The simulations may therefore
show a greater degree of sensitivity to perturbations in the deep interior.
Finally, to complete the description of the recipe, we apply the method of realization noise subtraction
\citep{hanasoge07a} in order to afford the ability to accurately study the effects of perturbations on the
wave travel times using short temporal simulation windows. Essentially, using identical realizations of
the source function, we perform two simulations: a `quiet' one with no perturbations and another with the
anomaly of choice. Because the calculation is linear and we use a linear method to extract the travel times 
\citep{gizon02}, the signal-to-noise ratio (SNR) of the measurement can be dramatically improved by subtracting
the travel times of the quiet data from the perturbed. See Figure~\ref{noise.sub} for a demonstration of this
procedure. As yet, this method is only possible in theory; there is no way of performing this sort of subtraction when
dealing with real data.

\subsection{Thermal Anomalies}
\begin{figure}[!ht]
\centering
\epsscale{1.0}
\vspace{-4.cm}
\plotone{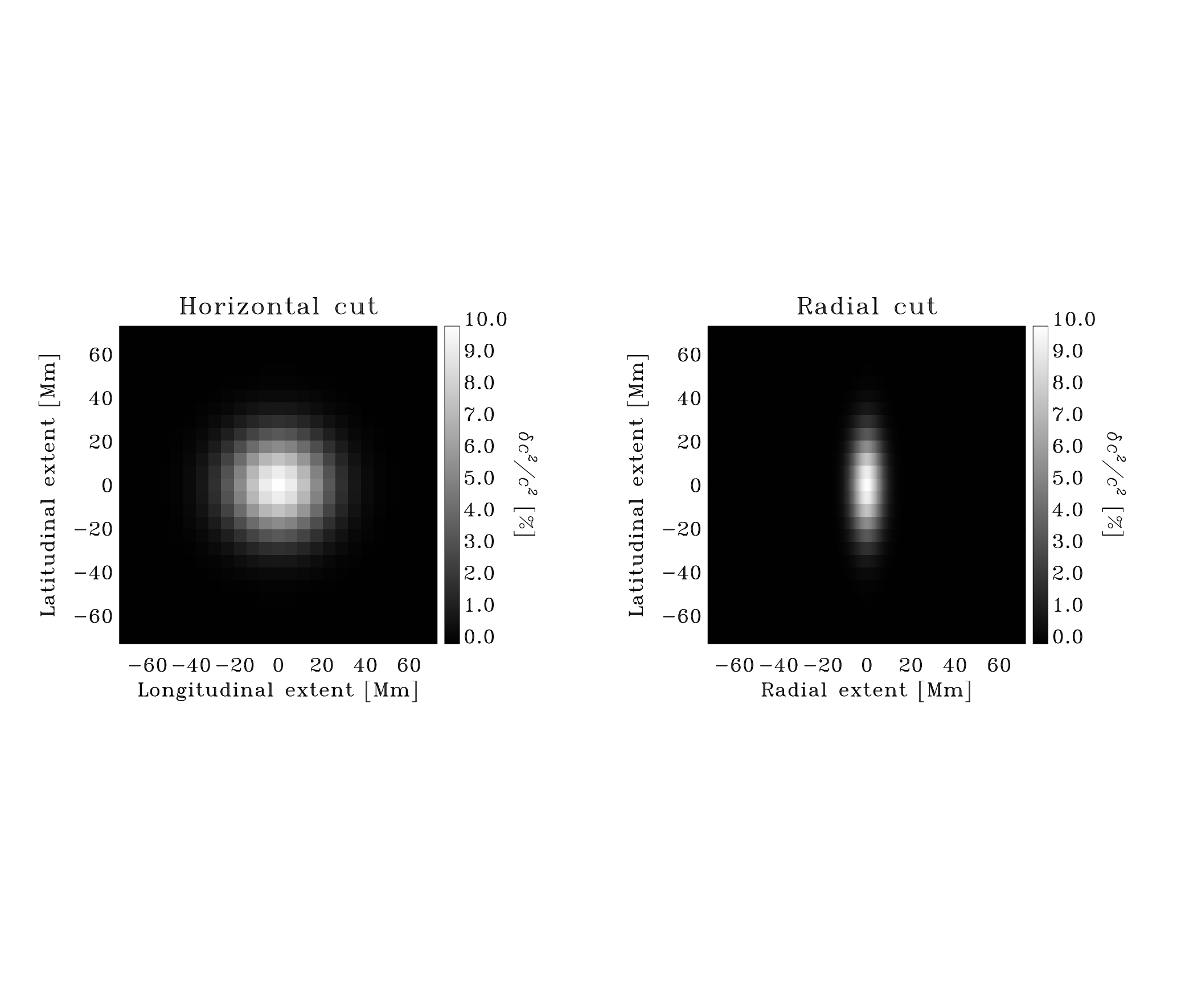}
\vspace{-4.5cm}
\caption{Spatial structure of the sound-speed perturbation. A horizontal and radial cut are displayed. Three cases
with the perturbation placed at different radial locations, $r=0.55, 0.7, 1.0 R_\odot$ are studied. In all cases,
the perturbation is placed at the equator.}
\label{pert_ss}
\end{figure}

The existence of a form of thermal asphericity in the tachocline has been suspected for a while now \citep[e.g.][]{jcd}.
We study here the possibilities relating to the inference of the nature of these deep-interior anomalies.
Changes in the thermal structure and hence the sound speed are effected by altering $\Gamma_1$, the first adiabatic index, as opposed to
perturbations in the pressure or density of the background model which result in hydrostatic inconsistencies
and consequently, strong numerical instabilities. Spatially, the perturbations resemble pancakes, thin in the 
radial direction and broad in the horizontal $(\theta,\phi)$ directions, where $(\theta,\phi)$ denote 
latitude and longitude respectively. In Figure~\ref{pert_ss}, we show the sound-speed perturbations.

\begin{figure}[!ht]
\centering
\epsscale{1.0}
\plottwo{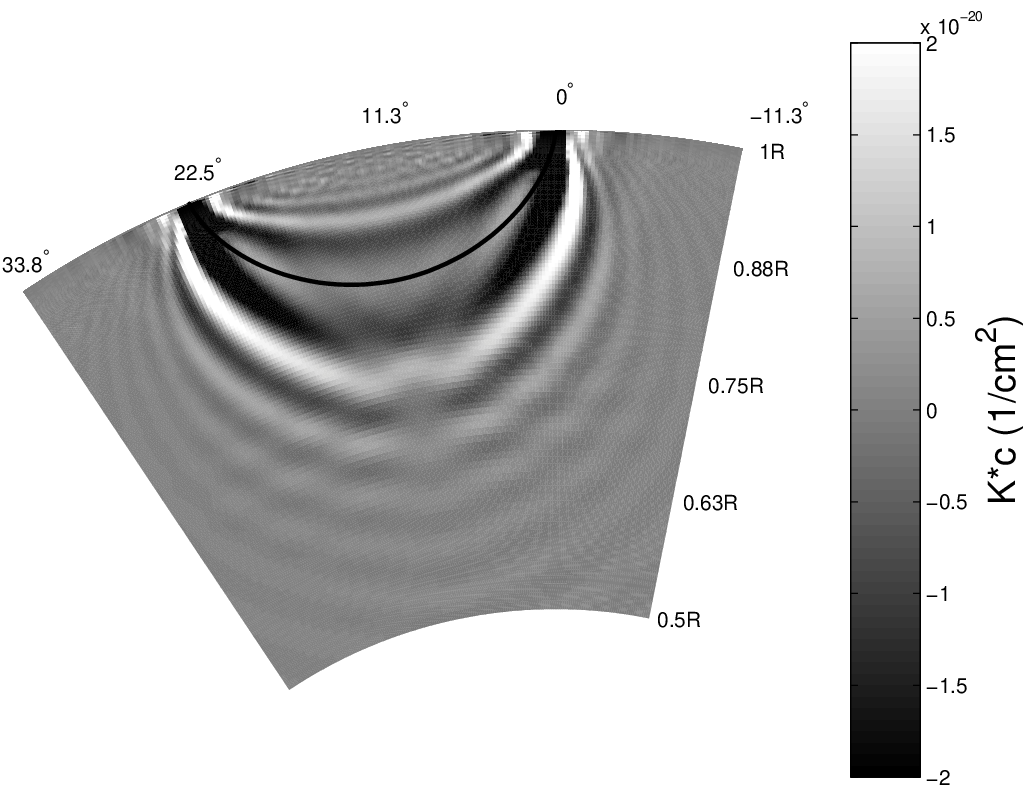}{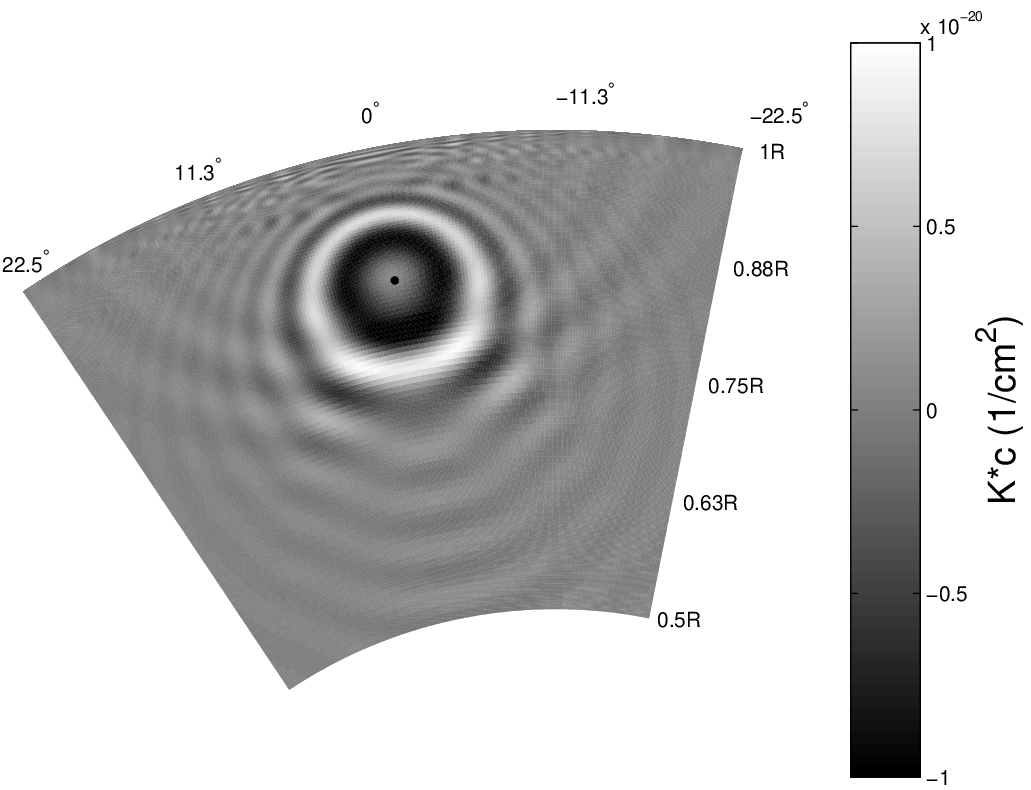}
\caption{Radial (left panel) and horizontal (right panel) slices of a sound-speed kernel, reproduced from \citet{birch00}.
A sensitivity kernel is defined as the response as seen in the travel times of a delta-function
sized perturbation. In this case, the kernel was computed for the monochromatic frequency of 5 mHz.
Complex interference patterns are seen in both slices - provided perturbations are sufficiently small,
the alternating signs of the lobes engender oscillating time shifts.}
\label{kernel_ss}
\end{figure}

\subsection{Time-distance analysis}\label{td.analysis.ss}
We use the common midpoint method of analysis \citep{duvall03} to average the data and extract $p$-mode travel 
times. Note that we do not perform any form of phase-speed filtering. The analysis proceeds as follows: for a
given point $(\theta_0, \phi_0)$ and a travel distance at the surface $\Delta$, we search for all points that lie on a circle
of diameter $\Delta$ with center $(\theta_0, \phi_0)$. Subsequently, we cross correlate signals at points that
are located diametrically opposite each other and average these pairwise cross correlations over the circular
region. We then fit the averaged cross correlation using the method of \citet{gizon02} to obtain the
travel time at the point $\tau(\theta_0,\phi_0)$. Since there is no directional bias in the averaging procedure,
the quantity $\tau$ is classified as the `mean travel time' and is sensitive primarily to sound-speed anomalies
along the ray path.

\begin{figure}[!ht]
\centering
\epsscale{1.0}
\vspace{-3.5cm}
\plotone{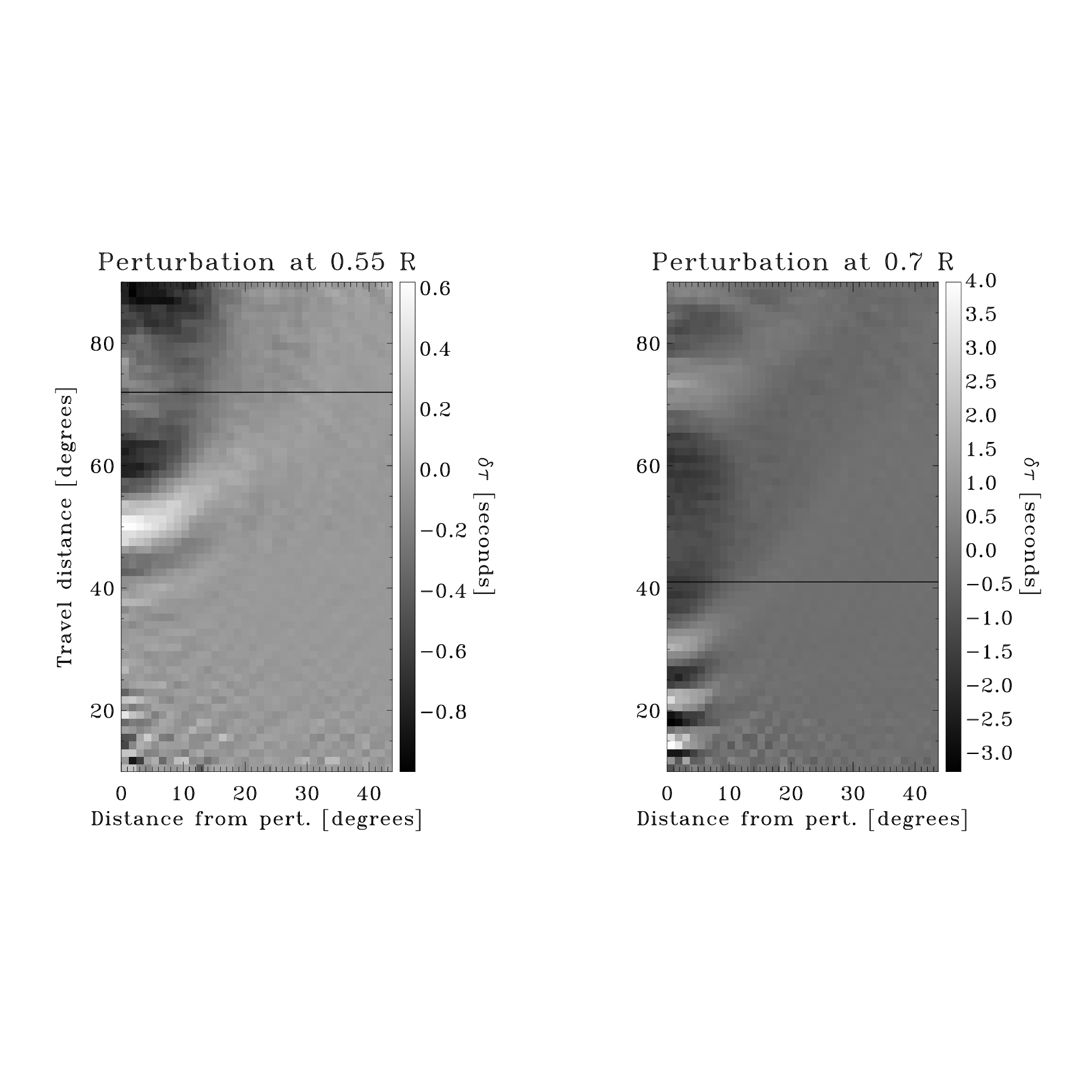}
\vspace{-4.5cm}
\caption{Time shifts caused by localized sound-speed perturbations (see Figure~\ref{pert_ss}) centered at 
$r= 0.55, 0.7 R_\odot$. These shifts are recorded at a series of wave travel distances and at numerous horizontal 
locations on the surface. Because the perturbation exhibits a horizontal rotational symmetry (see 
Figure~\ref{pert_ss}) about its center point, 
we azimuthally average the time shifts around this point. The vertical axis is the travel distance, also the 
distance between correlation points. The horizontal axis is the angular distance of the measurement point 
from the horizontal center of the perturbation $(0,0)$. Wavelengths of 4 mHz waves at these depths are approximately 
70 and 58 Mm respectively. The radial size of the perturbation is significantly smaller than the wavelength in 
either case, a manifestion of which are the variations of the time shifts with travel distance. The two horizontal 
lines show the ray theoretic travel distance for waves whose inner turning point coincides with the radial 
position of the perturbation. Because the horizontal size of the perturbation is comparable to the wavelength
in the $r = 0.7 R_\odot$ case, we do not see a flip in the sign of the time shift as we move horizontally away
from the perturbation. It is however visible in the $r=0.55 R_\odot$, where the sound-speed perturbation is
decidedly sub-wavelength in size.}
\label{tt.pert.ss}
\end{figure}

In Figure~\ref{tt.pert.ss}, we show the time shifts associated with the interactions of the waves with the sound-speed 
perturbations. It is our understanding that the oscillations seen in the maps are a consequence of the highly
sub-wavelength nature of these interactions. The fact that time shifts of both signs are seen, inspite of
a solely positive change in the sound speed, is indicative of subtler interference phenomena that are at play.
The sensitivity kernels of \citet{birch00} show the appearance of these `Fresnel zone'-like structures (see Figure
~\ref{kernel_ss}), due in part to the complex interference patterns of waves at constant frequency. \citet{duvall06} have experimentally
recovered the oscillatory travel-time features of these sensitivity kernels that occur when sub-wavelength 
sized perturbations are encountered.

In Figure~\ref{tt.pert.ss}, the solid lines in the two panels indicate the ray theoretic
travel distance for waves whose inner turning point coincides with the radial position of the perturbation. It is
seen that waves whose turning points lie well above (shorter travel distance waves) the perturbation seem equally
if not more sensitive (due to the proximity of the propagation regions of these waves to the surface) to the
perturbation. Evidently, the theoretical picture of waves being sensitive onlyÊto perturbations located above their inner turning
points is too simplistic to explain this phenomenon. Wave scattering in the
Sun is a predominantly linear phenomenon, especially in the deep interior, where the waves and convection are
presumably decoupled. Thus the scattering redistributes modal energies across a range of wave numbers, at a fixed frequency.
And because we do not phase speed filter this data, the cross correlation for a specific travel distance retains
a sensitivity to contributions from scattered waves at different wave numbers (and therefore different travel distances).

\begin{figure}[!ht]
\centering
\epsscale{0.5}
\plotone{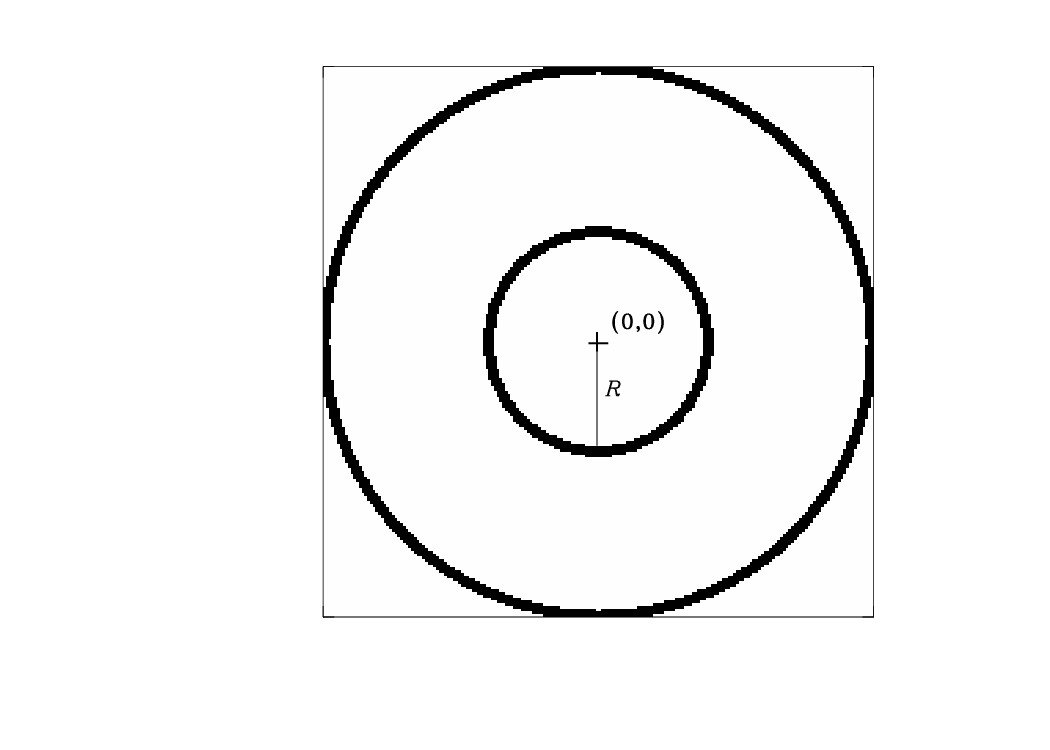}
\vspace{-1.cm}
\caption{Travel-time averaging geometry. The outer circle represents the observational boundaries of the sphere, while the inner indicates
points that are at distance $R$ from the location of the surface projection of the perturbation, $(0,0)$. Each point on the travel-time map is
associated with a travel time that is computed by (a) cross correlating signals at points at a distance $\Delta$ with it and (b) fitting the averaged
cross correlation by a Gabor wavelet. Within the noise 
level, we expect identical time shifts at all points along the inner circle
due to the (horizontal) directionally unbiased nature of wave scattering by the symmetric thermal anomalies of Figure~\ref{pert_ss}. Thus
we may average the time shifts on this annulus to improve the signal-to-noise properties of the analysis; we are then left with two
co-ordinates, the travel distance $\Delta$ and distance from the scatterer, $R$, as described in Figure~\ref{tt.pert.ss}.}
\label{ttgeom}
\end{figure}

The time shifts are measured according to the common mid-point procedure (described above) at a large number of 
points. The perturbation lies at horizontal co-ordinates $(0, 0)$
and as is evident from Figure~\ref{pert_ss}, rotationally symmetric about the center point. Thus,
azimuthally averaging the time shifts about $(0,0)$, we are left with two variables (see Figure~\ref{ttgeom}), namely the distance of the
point of measurement from the center of the perturbation (x-axis) and the travel distance (y-axis). Having therefore 
defined the geometry involved in the construction of Figure~\ref{tt.pert.ss},
we are ready to address questions relating to the fringes. Because the radial extent of the perturbation is
significantly smaller the wavelength of the acoustic waves at either depth, we must prepare ourselves for
the emergence of strong sub-wavelength effects as a function of the travel distance. And indeed, numerous fringes
appear in both cases as harbingers of the presence of this sub-wavelength feature. The kernel in Figure~\ref{kernel_ss}, which 
albeit is for a monochromatic wave, contains these oscillating positive and negative
signed lobes that hint at the possibility of such effects.

Meanwhile in the horizontal direction, we see some sign switching for the $r=0.55 R_\odot$ case while not so 
much for the other. This is because the horizontal size of the perturbation ($\sim 50$ Mm, see Figure~\ref{pert_ss})
is comparable to the wavelength of 58 Mm at $r=0.7 R_\odot$, while it is decidedly a subwavelength feature when compared
to the 70 Mm long waves at $r=0.55 R_\odot$. As one moves away from the perturbation, these alternating signed
lobes are encountered again (Figure~\ref{kernel_ss}). Depending on the size
of the perturbation, these lobes may end up being averaged, not showing up in the time shifts (size comparable to
or greater than the wavelength) or may result in oscillations (size smaller than the wavelength).

Although our knowledge of the structure and dynamics of the tachocline region is limited, it is certainly
well within the realm of possibility that there exist thermal asphericities whose dimensions may be
smaller than a wavelength. In such a situation, ray theory is untenable, and a wave mechanical 
description becomes necessary. A prominent source of solar observational data is the Michelson Doppler Imager 
\citep[MDI;][]{scherrer} instrument, onboard the Solar and Heliospheric Observatory (SOHO). The MDI medium-$l$
program has approximately 10-yr long line of sight Doppler velocity observations of the solar surface. However,
this data is somewhat corrupted by systematics such as strong center-to-limb variations ($\sim$ 6 -- 18s) in the zeroth order 
wave travel time, aliasing across the spatial Nyquist, fore-shortening, line-of-sight projection related effects etc. The resolution 
of these effects may be in sight \citep[e.g][]{zhao, duvall08} but the issues are outstanding as yet. Studying the 
interior thermal structure of the Sun requires the measurement of absolute quantities like the mean travel time, in contrast to 
investigations of flows which are described by relative quantities like travel-time differences. Absolute quantities are very 
difficult to pin down precisely because of this center-to-limb variation, sprouting questions like what value
of the mean travel time is `correct' and what part is systematic. For some of these reasons, we have not pursued
observational investigations of the thermal structure in the deep solar interior. We now turn to studies of
jets.

\section{Jets in the convection zone}\label{jets}
\begin{figure}[!ht]
\centering
\epsscale{1.0}
\vspace{-2.cm}
\plotone{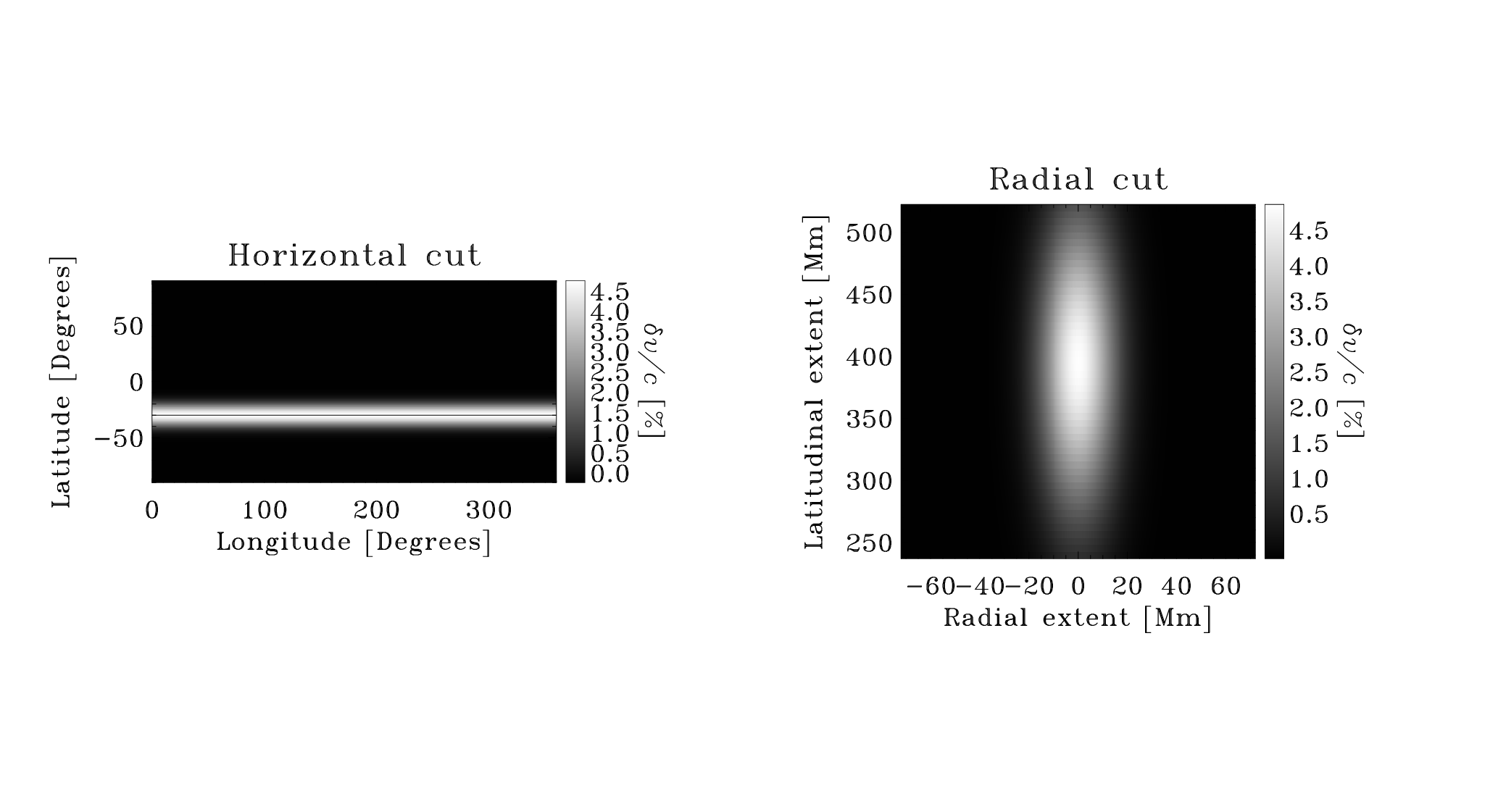}
\vspace{-2.cm}
\caption{Form of jets used in simulations. The jet structure is highly anisotropic, with large horizontal dimensions
but an extremely thin radial extent. The maximum jet velocity is set to 5\% of local sound speed, amounting to 11.25 km/s
and 8.5 km/s at $r=0.71, 0.81 R_\odot$ respectively. Although the model jet velocities may be unrealistically large, the time 
shifts may be linearly scaled with the flow velocity, thus allowing us to estimate the expected time shifts for
much weaker jets.}
\label{jet_pert}
\end{figure}

\begin{figure}[!ht]
\centering
\epsscale{0.4}
\plotone{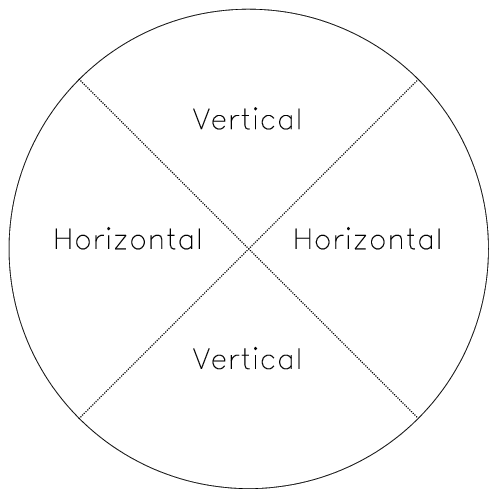}
\caption{Quadrant time-distance averaging geometry used for the study of flows. Points on each quadrant are cross
correlated with those diametrically opposite. The east-west quadrants are separated from the north-south so as
to study flows in these directions in isolation. For example, the difference time shift $\tau_{e-w} - \tau_{w-e}$ 
is an indicator of flow magnitude in the east west direction, where $\tau_{e-w}$ is the travel time from a point on
the east quadrant to its diametrically opposite counterpart on the west quadrant, and vice-versa for $\tau_{w-e}$. }
\label{quad}
\end{figure}

We choose jets with spatial structure of the form displayed in Figure~\ref{jet_pert}. We can no longer apply
the straightforward common midpoint analysis of $\S$\ref{td.analysis.ss} because of the inherent directional
bias that flows introduce. Waves that move along with flow are sped up and vice versa. Keeping the same geometry
as above, i.e. choosing a center point and searching for a set of points at a constant distance away from this point,
we then divide up the circle into four sectors: north, south, east, and west. Points on each quadrant are 
cross correlated with their diametrically opposite counterparts and the averaging is restricted to these
quadrants, thus allowing us to study north-south and east-west directed flows in isolation. See Figure~\ref{quad}
for an illustration of this geometry. Time shifts, as before, are computed for a variety of travel distances and 
at a large number of surface points. Because the jet is invariant over longitude, we average the time shifts
over all longitudes, thus leaving us two independent variables, namely the travel distance and latitude. 

\begin{figure}[!ht]
\centering
\epsscale{1.0}
\vspace{-2.cm}
\plotone{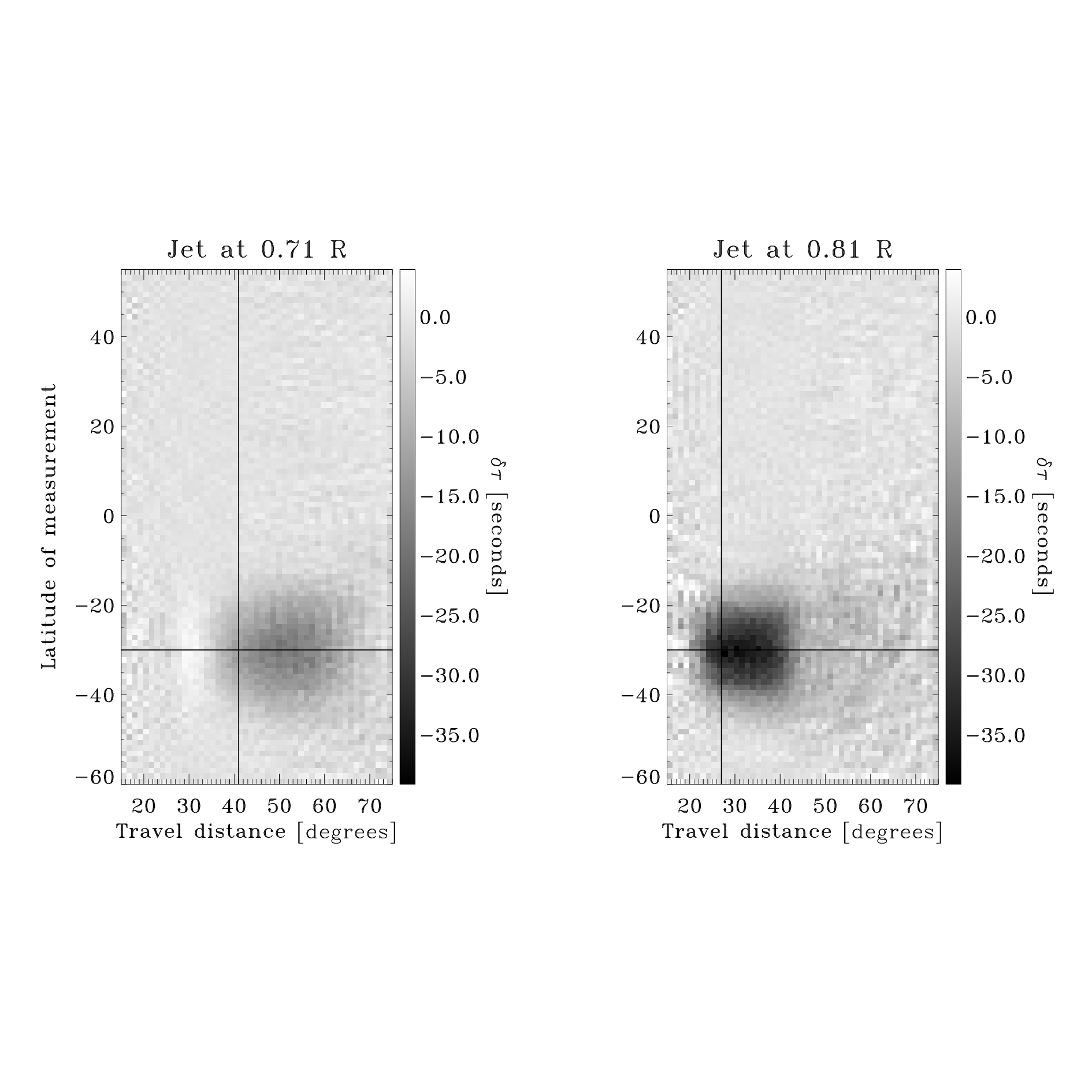}
\vspace{-4.cm}
\caption{Statistically significant difference time shifts caused by the interaction of waves with jets in the deep 
interior (see Figure~\ref{jet_pert}). Because the jet is axisymmetric, we average the difference time shifts over 
all longitudes. The cross-hairs show the center of the jet, i.e. the latitude about which the jet is centered 
(horizontal line) and the travel distance of a ray whose inner turning point is the radial center of the jet 
(vertical line). The horizontal extent of the jet is larger than a wavelength, resulting in time shifts of 
only one sign (as a function of latitude). Radially, it is sub-wavelength in size for the $r=0.71 R_\odot$ centered jet (wavelength 74 Mm), 
but comparable to the wavelength for the $r=0.81 R_\odot$ case (55 Mm long waves), thereby creating weak fringe-like 
patterns in the time shifts of the former but not so much the latter. The simulation was performed with $l_{max} = 95$ 
and hence the shortest travel distance that could be recovered from the data was approximately $15^\circ$.}
\label{tt.jets}
\end{figure}

The difference time shifts (so-called because we take the difference of the travel times) obtained with this
averaging scheme are shown in Figure~\ref{tt.jets}. Because of the large horizontal size of the jets, 
fringe-like time shift oscillations do not appear as the latitudinal distance from the jet grows. The radial
extent of jet ($\sim$ 40 Mm) is smaller than the wavelength at $r=0.71 R_\odot$ (58 Mm) but becomes comparable 
in the $r=0.81 R_\odot$ centered case (wavelength of 42 Mm). Thus the time shift fringes are seen in the former
but not in the latter. 

In the process of discovering these effects related to the sub-wavelength spatial dimensions of scatterers, it
was realized that by studying the amplitude of these oscillatory time shifts in comparison to the dominant 
shift, we could place bounds on the size of the scatterer. In other words, the tachocline thickness, thought
by some to be close to a wavelength \citep[e.g.][]{kosovichev96} and others to be much smaller
\citep[e.g.][]{gough99} is then a parameter that can be estimated by this method. Much to our chagrin however, 
the lack of a sufficient signal-to-noise ratio has blocked our efforts in this regard. This shall be the topic of discussion
for the remaining part of this paper.

\section{Study of the tachocline}\label{observations}
\begin{figure}[!ht]
\centering
\epsscale{1.}
\plotone{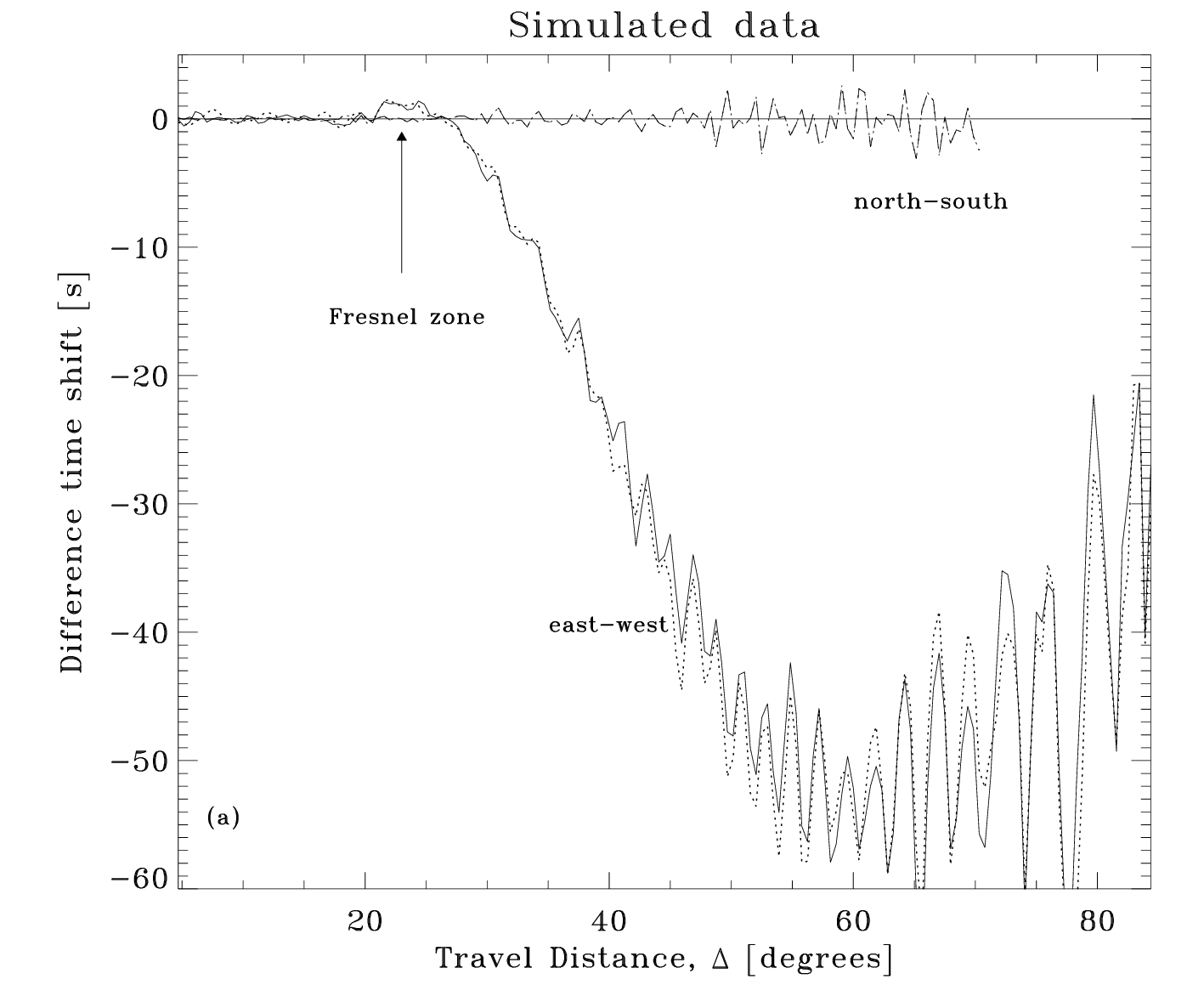}
\caption{Simulated difference time shifts averaged over the southern latitudes (dotted)
and the northern latitides (solid line). The simulated east-west time shifts contain a statistically significant Fresnel zone, seen for waves
that travel a distance of $25^\circ$ or so (the Fresnel zone disappears when the FWHM becomes comparable to a wavelength). 
The noise seen at larger distances is presumably from the poor quality of noise subtraction.}
\label{simulated_we_ns}
\end{figure}

Our initial assessment of the situation was that with a number of years of MDI medium-$l$ observations, a latitudinal band of
$\pm 30^\circ$ around the equator where differential rotation is nearly absent, we might have enough
data to conclude one way or the other about the thickness of the tachocline. Before analyzing observations,
we first theoretically investigated the effects of an angular velocity gradient at the base of the convection zone 
on the travel times. We performed simulations with a velocity perturbation that took the form of
 a rigidly rotating radiative interior (rotation rate of 9.065 $\mu{\rm Rad}~{\rm s}^{-1}$) tied to a non-rotating convection zone. 
The size of the interface between the rotating and non-rotating zones was varied in order to sharpen/weaken the gradient.
With sufficient sharpness, a sub-wavelength sized feature could be created and vice versa. In Figure~\ref{simulated_we_ns}
we display the time shifts recovered from this simulation (the rotation gradient at the interface was $1.1 \times 10^{-4} {\rm s}^{-1}$); 
a Fresnel zone is clearly observed. 

\begin{figure}[!ht]
\centering
\epsscale{0.6}
\plotone{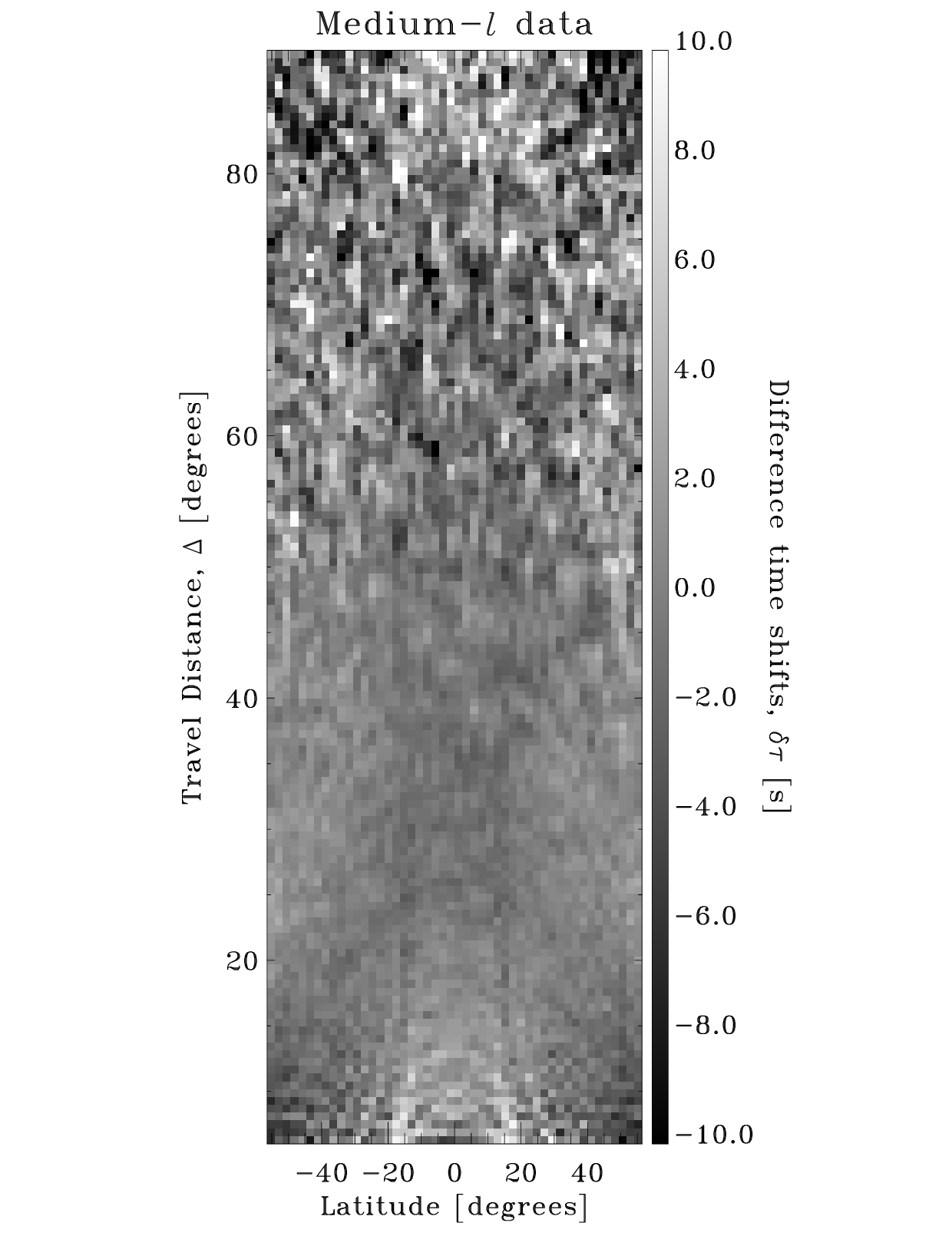}
\caption{Longitudinally averaged difference time shifts obtained from analyses of 1 year of untracked MDI medium-$l$ observations. Subtle 
effects such as Fresnel zones are entirely masked by the noise in the travel-time map. A zeroth order rotation related
time difference has been subtracted for each distance. This benchmark time was estimated by averaging cross covariances over the entire 
available disk and fitting the resultant. We believe that the features seen extending over the small distance range
are due to systematical issues of unknown origin.}
\label{plot.1}
\end{figure}

\begin{figure}[!ht]
\centering
\epsscale{0.6}
\plotone{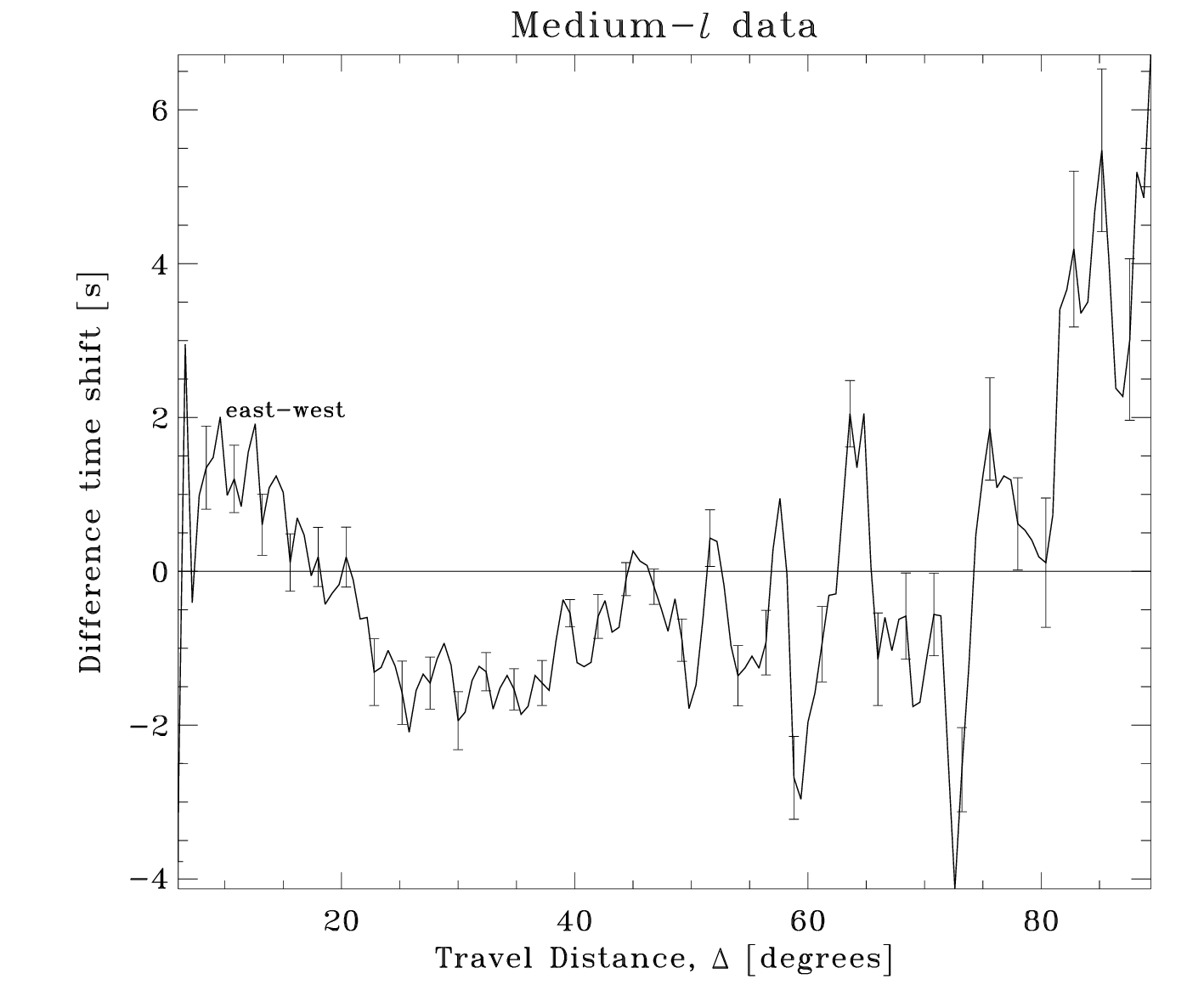}
\caption{Difference time shifts of Figure~\ref{plot.1} averaged over the entire latitude range. Fresnel zones, if any,
hide well beneath the noise of the travel-time map, which are indicated by the vertical bars.}
\label{data_we_ns}
\end{figure}

As a first step, we analyzed and extracted rotation signals from 1 year of medium-$l$ observations (see Figure~\ref{plot.1}).
The Fresnel zone signals are expected to scale linearly with the rotation gradient at the base of the convection zone. From the
rotation inversions of observed frequency splittings \citep[e.g.][]{schou}, we estimate the rotation gradient to be at least $1.44 \times 10^{-6} {\rm s}^{-1}$, two orders of
magnitude less than the simulated values. The Fresnel zone time shift associated with the solar rotation gradient is possibly 
0.01-0.1 seconds or less (the peak modal power occurs at a lower frequency in the Sun than the simulations) in magnitude. Two stumbling blocks 
lie on the path to the detection of solar Fresnel zones: (1) precisely determining
the zeroth order rotation related time shift curve, and (2) achieving high signal-to-noise ratios. While the former presents a somewhat formidable
challenge, the latter seems to lie in the realm of impossibility, at least with the current quantity of data available. The error bars in Figure~\ref{data_we_ns} 
for the one year period appear to be the order of 0.5-0.6 seconds; since the noise goes down as the square root of observational time, we expect a $\sqrt{10} = 3.16$
reduction, setting the error bars at around 0.15 - 0.2 seconds. Unfortunately this may well be much larger than the estimated observational effect.

\section{Conclusions}\label{conclude.sec}
We have attempted to explore the prospects for the seismology of the solar deep interior through the 
application of techniques of deep-focusing time-distance helioseismology on numerical
simulations and MDI observations. The solar dynamo is possibly closely tied to the dynamics in
the tachocline, a remarkably important but relatively poorly understood region. In this
regard, any one of the following would be very useful: strong observational constraints on 
the thickness of the tachocline, inferences of rotational deviants such as jets \citep{jcd05}, the structure of the thermal 
anomalies, etc. Azimuthally symmetric features are probably better off being studied using the
methods of global helioseismology; however, spatially limited perturbations are perhaps more 
amenable to the techniques of local helioseismology.

The study of the interactions between waves and sub-wavelength sized perturbations bring to
the fore a means of using wave statistical information. We have attempted to exploit the premise that
features with sub-wavelength spatial dimensions create Fresnel zones in the difference shifts 
(while the super-wavelength counterparts do not) in order to
place strong constraints on the angular velocity gradient at the bottom of the convection zone. 

The prospects? Figure~\ref{plot.1} presents an unexciting forecast for local helioseismic investigations
of the solar tachocline and the nearby regions. However, encouraging results are derived for studies of 
the moderately deep convection zone and near-surface regions, which provide reasonable signal-to-noise ratios.

\acknowledgements
The computing was performed on the NASA supercomputer Columbia, housed in the Ames research center. S. M. H. would
like to acknowledge funding from NASA grant HMI NAS5-02139.



\begin{thebibliography}{}
\bibitem[Birch \& Kosovichev(2000)]{birch00} Birch, A. C. \& Kosovichev, A. G. 2000, ApJ, 192, 193
\bibitem[Bogdan \& Cally(1995)]{bogdan} Bogdan, T. J., \& Cally, P. S. 1995, ApJ, 453, 919
\bibitem[Bogdan(1997)]{bogdan97} Bogdan, T. J. 1997, ApJ, 477, 475
\bibitem[Bozhevolnyi \& Vohnsen(1996)]{boz} Bozhevolnyi, S. I. \& Vohnsen, B. 1996, Physics Review Letters, 77, 3351
\bibitem[Christensen-Dalsgaard et al.(1995)]{jcd95} Christensen-Dalsgaard, J., Monteiro, M. J. P. F. G., \& Thompson, M. J.  1995, MNRAS, 276, 283 
\bibitem[Christensen-Dalsgaard et al.(1996)]{jcd} Christensen-Dalsgaard, J., et al. 1996, Science, 272, 1286 
\bibitem[Christensen-Dalsgaard(2002)]{jcd02} Christensen-Dalsgaard, J. 2002, RvMP, 74, 1073 
\bibitem[Christensen-Dalsgaard et al.(2003)]{jcd03} Christensen-Dalsgaard, J., Thompson, M. J., Miesch, M. S., \& Toomre, J. 2003, ARA\&A, 41, 599 
\bibitem[Christensen-Dalsgaard et al.(2005)]{jcd05} Christensen-Dalsgaard, J. et al. 2005, ASPC, 346, 115
\bibitem[Couvidat, Birch, \& Kosovichev(2006)]{couvidat} Couvidat, S., Birch, A. C., \& Kosovichev, A. G. 2006, ApJ, 640, 516
\bibitem[Duvall et al.(1993)]{duvall} Duvall, T. L., Jefferies, S. M., Harvey, J. W., \& Pomerantz, M. A. 1993, \nat, 362, 430
\bibitem[Duvall(2003)]{duvall03} Duvall, T. L., Jr. 2003, SOHO 12/GONG+ 2002 proceedings, Ed.: H. Sawaya-Lacoste, ESA Publications Division, 259
\bibitem[Duvall et al.(2006)]{duvall06} Duvall, T. L., Jr., Birch, A. C., \& Gizon, L. 2006, ApJ, 646, 553
\bibitem[Duvall et al.(2008)]{duvall08} Duvall, T. L., Jr., Hanasoge, S. M., \& Birch, A. C., {\it in prep}
\bibitem[Elliott \& Gough(1999)]{gough99} Elliott, J. R. \& Gough, D. O. 1999, ApJ, 516, 475
\bibitem[Gizon \& Birch(2002)]{gizon02} Gizon, L., \& Birch, A. C. 2002, ApJ, 571, 966
\bibitem[Gizon et al.(2006)]{gizon06} Gizon, L., Hanasoge, S. M., \& Birch, A. C. 2006, ApJ, 643, 549
\bibitem[Gizon \& Birch(2005)]{gizon05} Gizon, L. \& Birch, A. C. 2005, Living Reviews in Solar Physics, 2, 6
\bibitem[Hanasoge et al.(2006)]{hanasoge1} Hanasoge, S. M. et al. 2006, \apj, 648, 1268
\bibitem[Hanasoge et al.(2007a)]{hanasoge07a} Hanasoge, S. M., Duvall, T. L., Jr., \& Couvidat, S. 2007a, \apj, 664, 1234
\bibitem[Hanasoge(2007)]{hanasoge.phd} Hanasoge, S. M. 2007, Ph. D. thesis, Stanford University, http://soi.stanford.edu/papers/dissertations/hanasoge/
\bibitem[Hanasoge et al.(2008)]{hanasoge07b} Hanasoge, S. M., Birch, A. C., Bogdan, T. J., \& Gizon, L. 2008, \apj, 680, 774
\bibitem[Hu et al.(1996)]{hu} Hu, F. Q., Hussaini, M. Y., \& Manthey, J. L. 1996, Journal of Computational Physics, 124, 177
\bibitem[Kosovichev(1996)]{kosovichev96} Kosovichev, A. G. 1996, ApJ, 469L, 61
\bibitem[Kosovichev \& Duvall(1997)]{kosovichev} Kosovichev, A. G. \& Duvall, T. L., Jr. 1997, SCORe proceedings, Ed.: F.P. Pijpers, J. Christensen-Dalsgaard, and C.S. Rosenthal, Kluwer Academic Publishers, 241
\bibitem[Lele(1992)]{lele} Lele, S. K. 1992, Journal of Computational Physics, 103, 16
\bibitem[Maynard et al.(1985)]{maynard} Maynard J. D., Williams, E. G., \& Lee, Y. 1985, Journal of the Acoustical Society of America, 78, 1395
\bibitem[Scherrer et al.(1995)]{scherrer} Scherrer et al. 1995, \solphys, 162, 129
\bibitem[Schou et al.(1998)]{schou} Schou, J. et al. 1998, \apj, 505, 390
\bibitem[Thompson(1990)]{thompson} Thompson, K. W. 1990, Journal of Computational Physics, 89, 439
\bibitem[Zhao et al.(2007)]{zhao} Zhao, J., Hartlep, T., Kosovichev, A. G., \& Mansour, N. N. 2007, AGU abstract
\end{thebibliography}
\end{document}